\documentclass[aps,prl,twocolumn,superscriptaddress,showpacs,amsmath,amssymb]{revtex4}

\usepackage{graphicx}

\begin{document}

\title{Optical nanoprobing via spin-orbit interaction of light}

\author{Oscar G. Rodr\'iguez-Herrera}
\affiliation{Applied Optics Group, School of Physics, National University of Ireland, Galway, Galway, Ireland}

\author{David Lara}
\affiliation{The Blackett Laboratory, Imperial College London, SW7 2BW, London, United Kingdom}

\author{Konstantin Y. Bliokh}
\affiliation{Applied Optics Group, School of Physics, National University of Ireland, Galway, Galway, Ireland}

\author{Elena A. Ostrovskaya}
\affiliation{ARC Centre of Excellence for Quantum-Atom Optics and
Nonlinear Physics Centre, Research School of Physics and Engineering, The Australian National University,
Canberra ACT 0200, Australia}

\author{Chris Dainty}
\affiliation{Applied Optics Group, School of Physics, National University of Ireland, Galway, Galway, Ireland}


\begin{abstract}
We show, both theoretically and experimentally, that high-numerical-aperture (NA) optical microscopy is accompanied by strong spin-orbit interaction of light, which translates fine infomation about the specimen to the polarization degrees of freedom of light. An 80nm gold nano-particle scattering the light in the focus of a high-NA objective generates angular momentum conversion which is seen as a non-uniform polarization distribution at the exit pupil. We demonstrate remarkable sensitivity of the effect to the position of the nano-particle: Its subwavelength displacement produces the giant spin-Hall effect, i.e., macro-separation of spins in the outgoing light. This brings forth a far-field optical nanoprobing technique based on the spin-orbit interaction of light.
\end{abstract}


\maketitle

In the past few years \textit{spin-orbit interaction (SOI)} of light was intensively studied in connection with the spin-Hall effect in inhomogeneous media \cite{1,2,3,4,5} and conversion of the angular momentum (AM) of light upon focusing and scattering \cite{6,7,8,9,10}. These are a group of dynamical phenomena where the spin of light affects its orbital motion and vice versa. Effective far-field AM conversion arises naturally upon propagation of light through anisotropic media \cite{11,12,13}. The SOI in isotropic inhomogeneous media is a more fine, \textit{subwavelength} effect which exhibits intrinsic coupling between polarization and position of light. Therefore, the SOI phenomena are subtle in a paraxial field and become conspicuous only at subwavelength scales, namely, upon tight focusing of light  \cite{6,7,8,14,15} or scattering by small particles \cite{9,10}. Thus, \textit{the SOI carries fine information about light interaction with matter} and is highly highly attractive for nano-optical applications -- in particular, for for \textit{subwavelength probing and imaging}. However, observation of the SOI in the focused or scattered fields requires additional interaction with test particles or near-field measurements \cite{7,9,14,15}. The main obstacle in using direct far-field imaging of the SOI in non-paraxial light is that any collector lens produces its own SOI which distorts or even eliminates the initial effect \cite{14}. Thus, the fundamental challenge of practical importance is to translate the strong subwavelength-scale SOI to the far-field.

\begin{figure}[t]
\includegraphics[width=8.7cm, keepaspectratio]{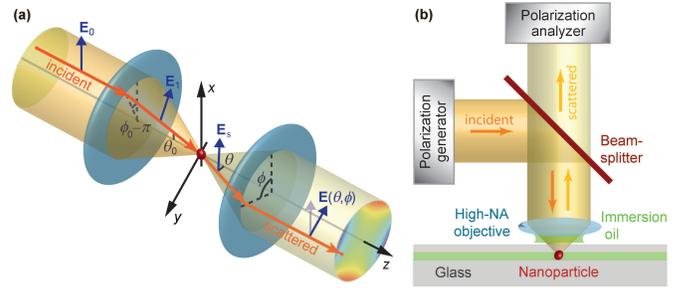}
\caption{(Color online) Schematics of the far-field imaging of a nano-particle, which involves the SOI of polarized light: (a) The ``lens-scatterer-lens'' system with the electric fields transformations and coordinates used in the text; (b) The experimental setup in the reflection configuration, where the focusing objective is also used as a collector lens and a beam splitter separates incident and scattered light.} \label{fig:schemes}
\end{figure}

In this Letter we show that \textit{a high-NA microscopy of nano-objects with polarized light} inherently involves effective far-field imaging of the SOI. A generic optical microscope consists of three basic elements: (i) high-NA focusing lens for the incoming paraxial light, (ii) a scattering specimen placed in the sensitive focal field, and (iii) high-NA lens capturing the scattered radiation in the far-field, Fig.~\ref{fig:schemes}(a). Owing to the joint action of all of these elements, the strong SOI produced by interaction of the focused field with the nano-particle is translated to \textit{an inhomogeneous polarization distribution of the outgoing paraxial field}. By analyzing the changes in the far-field polarization, it is possible to retrieve fine information about the scattering particle. Thus, by adding a polarization analyzer to the microscope, we obtain \textit{a sensitive angle-resolved far-field optical probe operating on the nano-scale via SOI of light}. As an illustration, we demonstrate clear polarimetric detection of subwavelength displacements of the particle.

The experimental setup that implements the system in Fig.~\ref{fig:schemes}(a), is shown schematically in Fig.~\ref{fig:schemes}(b). The incident polarization state is generated by the modulation of a linearly polarized laser beam ($\lambda=532$ nm) with a tandem of two Pockels cells. Next, the light is expanded to cover the entrance pupil of the high-NA objective (${\rm NA}=1.40$ with the refractive index of the immersion oil $n_{\rm oil} = 1.518$). The objective focuses the light, creating a non-paraxial 3D field that interacts with the specimen. We use a reflection configuration, where the same objective collects the scattered light, collimates it, and sends it to the spatially resolved polarization state analyzer. The latter consists of a division-of-amplitude polarimeter \cite{16} with four charge-coupled device (CCD) cameras that measure the Stokes parameters $\vec S = \left( {S_0 ,S_1 ,S_2 ,S_3 } \right)$ of the polarization state across the exit pupil of the objective:  $\vec S = \vec S\left( {x,y} \right)$.

An $80$nm diameter gold nano-sphere on a microscope cover glass was used as a specimen under oil immersion conditions 
(Fig.~\ref{fig:schemes}(b)). 
The position of the particle in the focal region was controlled using a three-axis piezoelectric positioning stage. 
The polarization distributions of the scattered light were measured for different input polarizations and at different positions of the nano-sphere, and then combined into the angle-resolved Mueller matrix for each location of the nano-sphere (details will be published elsewhere). This completely characterizes polarization properties of the system. The main limitation of our measurements is the presence of interference fringes from multiple reflections in the optical elements.

\begin{figure}[t]
\includegraphics[width=8.5cm, keepaspectratio]{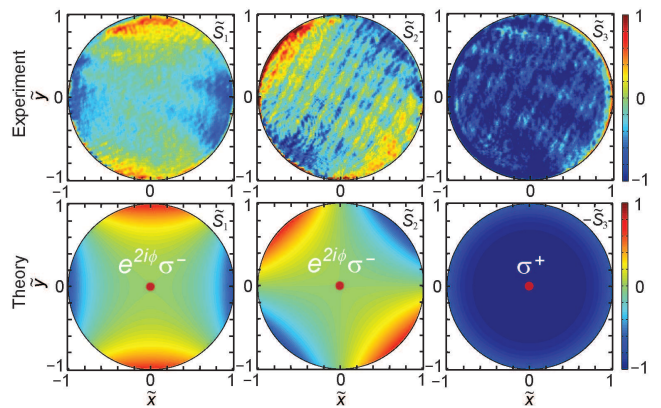}
 \caption{(Color online) Spin-to-orbital AM conversion of light when the nano-particle is located precisely in the focal point. Experimentally measured and theoretically calculated from Eq.~(4) spatial distributions of the normalized Stokes parameters $\tilde S_{1,2,3}  = S_{1,2,3} /S_0$ are shown for the case of circularly R-polarized incident light. Hereafter we use normalized coordinates $(\tilde x,\tilde y) = (x/d,y/d)$, $d = f{\mathop{\rm NA}\nolimits} /n_{\rm oil}$, and the sign difference in $S_3$ between the theory and experiment arises from the helicity flip in the reflection configuration.} \label{fig:focus}
\end{figure}

\begin{figure}[t]
\includegraphics[width=8.5cm, keepaspectratio]{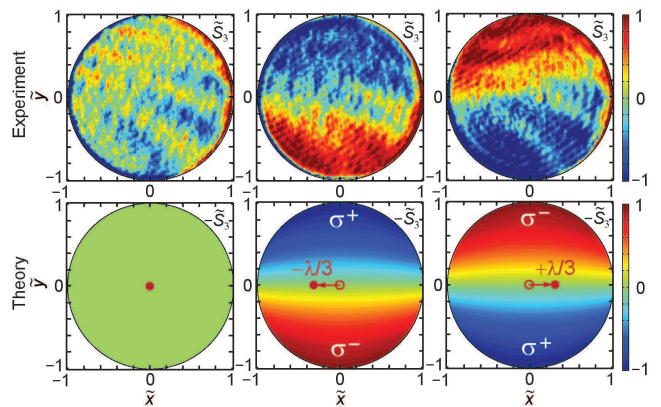}
\caption{(Color online) Giant spin Hall effect of light caused by subwavelength displacements of the nano-particle. Experimentally measured and theoretically calculated from Eq.~(\ref{eqn:transform}) spatial distributions of the SAM density  $\tilde S_3  = S_3 /S_0$ are shown for the case of linearly $x$-polarized incident light. Left-hand, middle, and right-hand panels correspond to the positions of the particle $x_{\rm s}  = 0$, $-\lambda/3$, and $\lambda/3$, respectively.} \label{fig:off}
\end{figure}

\begin{figure}[t]
\includegraphics[width=8.8cm, keepaspectratio]{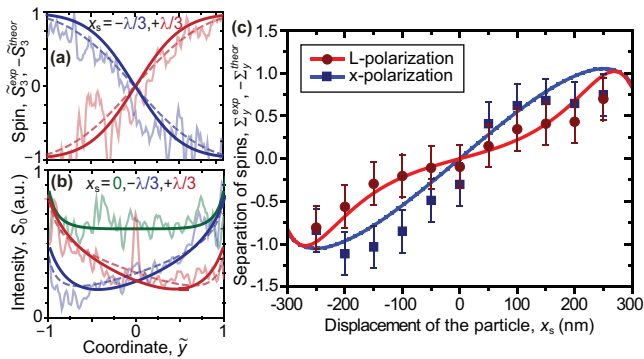}
\vspace{-0.6cm} \caption{(Color online) (a) Distributions of the SAM density ${\tilde S_3 \left( y \right)} |_{x = 0}$ at the particle displacements $x_{\rm s}  =  \pm \lambda /3$ for the linearly $x$-polarized incident light, cf. Fig.~\ref{fig:off}. Dark, dashed, and light curves represent theoretical calculations from Eq.~(\ref{eqn:transform}), approximate analytical calculations from Eqs.~(\ref{eqn:matrix}) and (\ref{eqn:correction}), and experimental measurements, respectively. (b) The spin Hall effect in the intensity distribution  ${S_0 \left( y \right)} |_{x = 0}$ at the particle positions $x_{\rm s}  = 0$, $ \pm \lambda /3$ for the R-polarized incident light. (c) Theoretically calculated from Eq.~(\ref{eqn:transform}) (curves) and experimentally measured (symbols) integral measure of the separation of the SAM, $\Sigma _y$, vs. position of the particle, $x_{\rm s}$, for the $x$- and L-polarized incident light.}
\label{fig:curves}
\end{figure}

The experimental results presented in Figures~\ref{fig:focus} and \ref{fig:off} clearly show that the output polarization exhibits non-uniform patterns. Namely, when the scatterer is located precisely at the focal point and illuminated by circularly-polarized light, the Stokes parameters  $S_1$
and $S_2$ possess four-fold symmetric patterns (Fig. \ref{fig:focus}) typical for the \textit{spin-to-orbit AM conversion} of light upon scattering \cite{13,17}. At the same time, the $S_3$ component, that characterizes the local density of the spin AM (SAM), is rather uniform. Strikingly, the sub-diffraction-limit displacement of the particle by the distance $\lambda/3$ along the $x$-axis induces dramatic changes in the output polarization patterns, breaking  their symmetry along the orthogonal $y$-axis and thus signaling the SOI of light in the system. Figure~\ref{fig:off} shows drastic displacement-induced separation of the two spin states (positive and negative $S_3$) in the linearly $x$-polarized light, i.e., the \textit{giant spin-Hall effect of light}. Instead of a tiny shift of the light position caused by switching between different spin states, which is typical for usual spin-Hall effect \cite{2,3,4,5,6,14,15}, here we observe \textit{macroscopic} redistribution of spins caused by the \textit{subwavelength} displacement of the scatterer. This effect can also be seen in the spin-dependent output intensity redistribution of the circularly-polarized light (Fig. \ref{fig:curves}(b)) \cite{remark}.

To describe the polarization transformations produced by the SOI in the system Fig.~1, we first employ the Debye--Wolf approach to analyze the effect of the high-NA lens \cite{18,19}. It assumes that the transverse wave electric field is transported along each ray refracted by the lens without change of the polarization state in the ray-accompanying coordinate frame. This resembles adiabatic approximation underlying the spin-redirection geometric phase \cite{21}. 
Hence, the 3D evolution of the field along the partial ray consists of a series of rotations \cite{19}: ${\bf{E}}_1 \left( {\theta _0 ,\phi _0 } \right) = \hat T_{{\mathop{\rm lens}\nolimits} } \left( {\theta _0 ,\phi _0 } \right){\bf{E}}_0$,
\begin{equation}\label{eqn:rotations}
\hat T_{{\mathop{\rm lens}\nolimits} }  = \sqrt {\cos \theta _0 } \,\hat R_z \left( { - \phi _0 } \right)\hat R_y \left( {- \theta _0 } \right)\hat R_z \left( {\phi _0 } \right)~.
\end{equation}
Here direction of the refracted ray is given by spherical angles  $\left( {\theta _0 ,\phi _0 } \right)$,   ${\bf E}_0$ (${\bf E}_1$) is the electric fields before (after) the lens (Fig.~\ref{fig:schemes}(a)),  $\hat R_n \left( \gamma  \right)$ is the operator of the rotation of the coordinate frame about the  $n$-axis by the angle $\gamma$, and  $\sqrt{\cos \theta _0}$ is the apodization factor \cite{18}.

Equation~(\ref{eqn:rotations}) describes the spin-to-orbit AM conversion of light upon focusing \cite{6,7,8,9,10}, revealing its purely \textit{geometrical} origin. The three successive rotations, $\hat R_z \left( {\phi _0 } \right)$, $\hat R_y \left( {- \theta _0 } \right)$, and  $\hat R_z \left( { - \phi _0 } \right)$, indicate, respectively: (i) the azimuthal rotation superimposing the  $xz$-plane with the local meridional plane, (ii) refraction on the angle $\theta _0$ therein, and (iii) the reverse azimuthal rotation compensating the first one (Fig.~\ref{fig:schemes}(a)). Because of the non-commutativity of the rotations, the changes in the projection of the electric field onto the laboratory  $xz$-plane are accompanied by azimuthal geometrical phases. Let us consider the spin eigenstate $\left| {\sigma ^ +  } \right\rangle$, or the right-hand circularly (R-) polarized light, carrying SAM $\sigma ^ +   = 1$ per photon (in units of  $\hbar$). The transformation~(\ref{eqn:rotations}) and projection onto the  $xy$-plane (which \textit{squeezes} polarization circle into ellipse) yield: $\left| {\sigma ^ +  } \right\rangle  \to a_0 \left| {\sigma ^ +  } \right\rangle  - b_0 \left| {\sigma ^ -  } \right\rangle e^{2i\phi _0 }$, $a_0  = \cos ^2 \left( {\theta _0 /2} \right)$, $b_0  = \sin ^2 \left( {\theta _0 /2} \right)$. 
This evidences partial conversion to the left-hand (L-) polarized state $\left| {\sigma ^ -  } \right\rangle$, which bears the SAM $\sigma ^ -   =  - 1$ per photon, with the \textit{helical geometric phase} $e^{2i\phi _0 }$. The latter signifies generation of the orbital AM (OAM) $l^ +   = 2$  per photon \cite{22} and ensures conservation of the total AM per photon: $\sigma ^ -   + l^ +   = \sigma ^ +$.
 
Next, we model the scattering of light on the nano-particle by a spherical wave generated by the electric dipole moment induced by the incident field \cite{20}. For a nano-particle located at ${\bf{r}} = {\bf{r}}_{\rm s}$ near the focal point ($r_{\rm s}   \ll \sqrt {f/k}$, where $k$  is the wave number and $f$  is the focal distance), the scattered electric field is
\begin{equation}\label{eqn:2}
{\bf{E}}_{\rm s}  \left( {\theta ,\phi } \right) \propto  -F \int  d^2 {\bf r}_0 F_0{\bar {\bf r} } \times \left[ {\bar {\bf r}} \times {\bf{E}}_1 \left( {\theta _0 ,\phi _0 } \right) \right]~,
\end{equation}
where $\bar{\bf r} = {\bf{r}}/r$ is the unit radial vector with spherical coordinates $\left( {\theta ,\phi } \right)$  of the observation point (Fig.~\ref{fig:schemes}(a)) and we performed the integration over all incoming fields:  $\int d^2 {\bf r}_0 \equiv \int\limits_0^{2\pi } {d\phi _0 \int\limits_0^\alpha  {d\theta _0 \sin \theta _0 } } $, with the aperture angle $\alpha  = {\rm sin}^{-1} \left({\rm NA}/n_{\rm oil}\right)$. The phase factors $F_0=e^{i\Phi _0 \left( {\theta _0 ,\phi _0 ,{\bf{r}}_{\rm s} } \right)}$ and $F= e^{i\Phi \left( {\theta ,\phi ,{\bf{r}}_{\rm s} } \right)}$ in Eq.~(2) account for deviations of optical paths between the incoming point ${\bf r}_0$, scatterer  ${\bf r}_{\rm s}$, and observation point  ${\bf r}$:  $\Phi _0  \simeq  - k{\bf{r}}_{\rm s} {\bf{r}}_0 /f$ and $\Phi  \simeq  - k{\bf{r}}_{\rm s} {\bf{r}}/f$. Because of the spherical symmetry, the transformation of the field upon scattering is quite similar to that upon focusing \cite{12,13}. Akin to Eq.~(\ref{eqn:rotations}), the vector operator in Eq.~(2) also consists of 3D rotations: $- \bar {\bf r} \times \left(  \bar {\bf r} \times  \right) = \hat R_z \left( -\phi  \right) \hat R_y \left( -\theta  \right) \hat P_z \hat R_y \left( \theta  \right) \hat R_z \left(\phi  \right)$ ($\hat P_z$ is the projector onto the $xy$-plane), and exhibits the spin-to-orbit AM conversion.

The whole system ``lens-scatterer-lens'', Fig.~\ref{fig:schemes}(a), is described by the combined action
of the focusing operator~(\ref{eqn:rotations}), the scattering transformation~(\ref{eqn:2}), and the operator of the collector lens $\hat T_{{\mathop{\rm lens}\nolimits} }^{ - 1} \left( {\theta ,\phi } \right)$. Since \textit{both input and output fields are paraxial}, the resulting transformation is reduced to the effective $2\times2$  \textit{angle-resolved Jones matrix} acting onto transverse  $xy$-components of the field: ${\bf{E}}_ \bot  \left( {\theta ,\phi ,{\bf{r}}_{\rm s} } \right) = \hat T_ \bot  \left( {\theta ,\phi ,{\bf{r}}_{\rm s} } \right){\bf{E}}_{0 \bot } $. Here
\begin{equation}\label{eqn:transform}
\hat T = F\,\hat T_{\rm lens}^{ - 1} \left( {\theta ,\phi } \right)\int  d^2 {\bf r}_0\,F_0\,\hat T_{\rm lens} \left( {\theta _0 ,\phi _0 } \right)
\end{equation}
is the 3D transformation and the subscript  $\bot$ denotes the projection onto the  $xy$-subspace.
The Jones matrix $\hat T_ \bot  \left( {\theta ,\phi ,{\bf{r}}_{\rm s} } \right)$ encapsulates SOI and polarization properties of the system; it describes complex output polarization patterns sensitive to the position of the particle. 

When the scatterer is located precisely at the focus, ${\bf{r}}_{\mathop{\rm s}\nolimits} = 0$, the matrix (\ref{eqn:transform}) can be calculated analytically. In the spin basis of circular polarizations this yields:
\begin{equation}\label{eqn:matrix}
\hat T_ \bot ^{(0)}  = \frac{A}{{\sqrt {\cos \theta } }}\left( {\begin{array}{*{20}c}
   a & { - be^{ - 2i\phi } }  \\
   { - be^{2i\phi } } & a  \\
\end{array}} \right)~,
\end{equation}
where $a = \cos ^2 \left( {\theta /2} \right)$, $b = \sin ^2 \left( {\theta /2} \right)$, and $A= (2\pi/15)\left[ {8 - \left( {\cos \alpha } \right)^{3/2} \left( {5 + 3\cos \alpha } \right)} \right]$. Non-diagonal elements of the Jones matrix (\ref{eqn:matrix}) signify effective \textit{spin-to-orbit AM conversion} in the paraxial field, which can be represented in the generic case as follows:
\begin{equation}\label{eqn:am}
\left| {\sigma _0 ,l_0 } \right\rangle  \to a\left| {\sigma _0 ,l_0 } \right\rangle  - b\left| {\sigma _0  \mp 2,l_0  \pm 2} \right\rangle~.
\end{equation}
Here the two quantum numbers denote the AM eigenstates of paraxial light: the SAM  $\sigma _0  =  \pm 1$ corresponds to the R and L modes, whereas the OAM $l_0  = 0, \pm 1, \pm 2,...$ corresponds to the helical phases $e^{il_0 \phi }$ \cite{22}. Figure \ref{fig:focus} shows the distributions of the output Stokes parameters calculated from the Jones matrix (\ref{eqn:matrix}) for the R input mode  $\left| {1,0} \right\rangle$. They clearly reproduce the four-fold symmetric patterns observed in the experiment and caused by the  $\left| { - 1,2} \right\rangle$ vortex component in the output field. The Jones matrix (\ref{eqn:matrix}) is typical for the AM conversion on \textit{anisotropic paraxial-field optical elements} \cite{11,12}, such as q-plates. However, in our case the AM conversion is based on the \textit{interaction of the non-paraxial focused field inside the system with the nano-scatterer}.

This fact leads to a remarkable sensitivity of the polarization properties of the system to the position of the nano-particle. If  ${\bf{r}}_{\mathop{\rm s}\nolimits}   \ne 0$, the first-order correction to the Jones matrix (\ref{eqn:matrix}) can be calculated analytically from the general operator (\ref{eqn:transform}) in the limit of small subwavelength displacements:   $kr_{\rm s}   \ll 1$. This results in the Jones matrix  $\hat T_ \bot   \simeq \hat T_ \bot ^{(0)}  - \hat T_ \bot ^{(1)} $ with
\begin{equation}\label{eqn:correction}
\hat T_ \bot ^{(1)}  =  i\frac{{B\sin \theta }}{{2\sqrt {\cos \theta } }}\left( {\begin{array}{*{20}c}
   {\rho_{\rm s} e^{ - i\phi } } & {\rho _{\mathop{\rm s}\nolimits} ^ *  e^{ - i\phi } }  \\
   {\rho_{\rm s} e^{i\phi } } & {\rho _{\mathop{\rm s}\nolimits} ^ *  e^{i\phi } }  \\
\end{array}} \right) + i\frac{C}{A}\zeta_{\rm s} \hat T^{(0)}~, 
\end{equation}
where $\rho_{\rm s}  = k(x_{\rm s}  + iy_{\rm s} )$, $\zeta_{\rm s}=k z_{\rm s}$, $B=(\pi/21)\left[ {8 - \left( {\cos \alpha } \right)^{3/2} \left( {11 - 3\cos 2\alpha } \right)} \right]$, and $C=(\pi/35)\left[ {12 - \left( {\cos \alpha } \right)^{5/2} \left( {7 + 5\cos \alpha } \right)} \right]$. Equation~(\ref{eqn:correction}) unveils \textit{SOI between polarization of the outgoing light and position of the paprticle}, $\rho_{\rm s}$, which manifests itself as an additional AM conversion in the system. Namely, upon transverse displacement of the particle, some parts of the SAM states $\left| {\sigma ^ \pm  } \right\rangle$ acquire vortices $e^{ \mp i\phi }$ bearing OAM $\mp 1$  per photon. This is because a transverse shift of the centre of gravity of the scattered field components $\left| {1,0} \right\rangle$ and $\left| { - 1,2} \right\rangle$ is accompanied by generation of a vortex component with the OAM differing by $1$ from the original field \cite{23}, resulting in $\left| {1, - 1} \right\rangle$ and $\left| { - 1,1} \right\rangle$ components, respectively. We have also verified the vortex and AM behavior described by Eqs.~(4) and (6) by \textit{ab initio} numerical simulations of the system; the results will be published elsewhere.

Figures 3 and 4 show that displacement-induced SOI, Eq.~(6), describes various manifestations of \textit{the giant spin Hall effect of light} in the polarization and intensity distributions. The theoretical and experimental results are in good agreement, and analytical approximation (\ref{eqn:correction}) works well even at relatively high displacements $kr_{\mathop{\rm s}\nolimits} \simeq 2$. The observed macroscopic spin Hall effect can be employed as quantitative measures of the position of the particle. For example, the separation of spins along the  $y$-axis can be characterized by the integral quantity  
$\Sigma _y  = \left\langle {\tilde y S_3 /S_0 } \right\rangle  \equiv \int_{ - 1}^1 {\tilde y {\tilde S_3 } |_{x = 0} } d\tilde y$. Figure \ref{fig:curves}(c) shows theoretically calculated dependence $\Sigma _y \left( {x_{\rm s} } \right)$
 vs. the results of the experimental measurements. By measuring this quantity, we are able to detect displacements of the particle up to $\lambda/10$ even in our rather noisy system. Improving the quality of the system, the accuracy can be increased significantly since a $\lambda/2$ displacement of the particle translates to a unit variation in $\Sigma _y$ (Fig.~\ref{fig:curves}(c)). This is a proof-of-principle demonstration of capability to detect subwavelength-scale properties of the particles by using SOI of light.

To summarize, we have shown that far-field imaging of nano-objects with polarized light is accompanied by strong SOI and AM conversion. These effects are produced by interaction of the tightly focused field with the specimen, as well as by the imaging system itself. Owing to their subwavelength geometric origin, the SOI phenomena are highly sensitive to the position and scattering properties of the nano-object. In particular, we have demonstrated the giant spin Hall effect produced by subwavelength displacements of a nano-particle. In general, far-field SOI polarimetry offers a new type of probing, whereby polarization degrees of freedom of light carry and reveal fine information about subwavelength objects.

This work was supported by Shimadzu Corporation, CONACYT, the European Commission (Marie Curie fellowiship), Science Foundation Ireland (grant 07/IN.1/I906), and the Australian Research Council.


\begin{thebibliography}{99}
\bibitem{1} V.S. Liberman and B.Y. Zel'dovich, Phys. Rev. A {\bf 46}, 5199 (1992).
\bibitem{2} M. Onoda, S. Murakami, and N. Nagaosa, Phys. Rev. Lett. {\bf 93}, 083901 (2004); K.Y. Bliokh and Y.P. Bliokh, Phys Lett. A {\bf 333}, 181 (2004); K.Y. Bliokh and Y.P. Bliokh, Phys. Rev. Lett. {\bf 96}, 073903 (2006).
\bibitem{3} O. Hosten and P. Kwiat, Science {\bf 319}, 787 (2008).
\bibitem{4} K.Y. Bliokh {\em et al.}, Nature Photon. {\bf 2}, 748 (2008).
\bibitem{5} A. Aiello {\em et al.}, Phys. Rev. Lett. {\bf 103}, 100401 (2009).
\bibitem{6} Z. Bomzon, M. Gu, and J. Shamir, Appl. Phys. Lett. {\bf 89}, 241104 (2006).
\bibitem{7} Y. Zhao {\em et al.} Phys. Rev. Lett. {\bf 99}, 073901 (2007).
\bibitem{8} T.A. Nieminen {\em et al.}, J. Opt. A: Pure Appl. Opt. {\bf 10}, 115005 (2008);
P.B. Monteiro {\em et al.}, Phys. Rev. A {\bf 79}, 033830 (2009);
A.Y. Bekshaev, Cent. Eur. J. Phys., DOI: 10.2478/s11534-010-0011-2 [arXiv:0910.0129].
\bibitem{9} C. Schwartz and A. Dogariu, Opt. Express {\bf 14}, 8425 (2006);
D. Haefner, S. Sukhov, and A. Dogariu, Phys. Rev. Lett. {\bf 102}, 123903 (2009).
\bibitem{10} L.T. Vuong {\em et al.}, Phys. Rev. Lett. {\bf 104}, 083903 (2010).
\bibitem{11} M.V. Berry, M.R. Jeffrey, and M. Mansuripur, J. Opt. A: Pure Appl. Opt. {\bf 7}, 685 (2005); E. Brasselet {\em et al.}, Opt. Lett. {\bf 34}, 1021 (2009).
\bibitem{12} E. Hasman {\em et al.}, Prog. Opt. {\bf 47}, 215 (2005); L. Marrucci, C. Manzo, and D. Paparo, Phys. Rev. Lett. {\bf 96}, 163905 (2006);
G.F. Calvo and A. Picon, Opt. Lett. 32, 838 (2007).
\bibitem{13} E. Brasselet {\em et al.}, Phys. Rev. Lett. {\bf 103}, 103903 (2009).
\bibitem{14} N.B. Baranova {\em et al.}, JETP Lett. {\bf 59}, 232 (1994); B.Y. Zel'dovich {\em et al.}, JETP Lett. {\bf 59}, 766 (1994).
\bibitem{15} K.Y. Bliokh {\em et al.}, Phys. Rev. Lett. {\bf 101}, 030404 (2008); Y. Gorodetski {\em et al.}, Phys. Rev. Lett. {\bf 101}, 043903 (2008).
\bibitem{16} C. Brosseau, {\em Polarized light} (Wiley, New York, 1998).
\bibitem{17} C. Schwartz and A. Dogariu, Opt. Lett. {\bf 31}, 1121 (2006).
\bibitem{remark} A \textit{vortex}-dependent displacement-induced transverse intensity redistribution (orbital Hall effect) was observed recently on larger scales of Mie scattering: V. Garbin {\em et al.}, New J. Phys. {\bf 11}, 013046 (2009), cf. K.Y. Bliokh and A.S. Desyatnikov, Phys. Rev. A {\bf 79}, 011807(R) (2009).
\bibitem{18} B. Richards and E. Wolf, Proc. R. Soc. London A {\bf 253}, 358 (1959).
\bibitem{19} P. T\"or\"ok, P.D. Higdon, and T. Wilson, Opt. Commun. {\bf 148}, 300 (1998).
\bibitem{20} M.V. Berry, Nature {\bf 326}, 277 (1987).
\bibitem{21} J.D. Jackson, {\em Classical Electrodynamics} (Wiley, New York, 1975).
\bibitem{22} L. Allen, M.J. Padgett, and M. Babiker, Prog. Opt. {\bf 39}, 291 (1999).
\bibitem{23} G. Molina-Terriza, J.P. Torres, and L. Torner, Phys. Rev. Lett. {\bf 88}, 013601 (2002).

\end{thebibliography}
\end{document}